\documentclass[authorversion, screen]{acmart}
\AtBeginDocument{%
  }

\setcopyright{rightsretained}
\acmDOI{}
\acmConference[CHI WS '25]{Affective Interaction \& Affective Computing - Past, Present \& Future - CHI Workshop 2025}{April 27,
  2025}{Yokohama, Japan}

\begin{document}

\title{Improving Inclusivity for Emotion Recognition Based on Face Tracking}

\author{Mats Ole Ellenberg}
\authornote{also with the Centre for Tactile Internet with Human-in-the-Loop (CeTI)}
\email{mellenberg@acm.org}
\orcid{0000-0002-0354-1537}
\affiliation{%
  \institution{Interactive Media Lab Dresden, \\ TUD Dresden University of Technology}
  \city{Dresden}
  \country{Germany}
}
\author{Katja Krug}
\authornotemark[1]
\email{katjakrug@acm.org}
\orcid{0000-0003-4800-6287}
\affiliation{%
  \institution{Interactive Media Lab Dresden, \\ TUD Dresden University of Technology}
  \city{Dresden}
  \country{Germany}
}

\renewcommand{\shortauthors}{Ellenberg et al.}

\begin{abstract}
The limited expressiveness of virtual user representations in Mixed Reality and Virtual Reality can inhibit an integral part of communication: emotional expression. Emotion recognition based on face tracking is often used to compensate for this. However, emotional facial expressions are highly individual, which is why many approaches have difficulties recognizing unique variations of emotional expressions. We propose several strategies to improve face tracking systems for emotion recognition with and without user intervention for the Affective Interaction Workshop at CHI '25~\cite{ahmadpour2025affective}. 
\end{abstract}



\maketitle

\section{Introduction and Background}





Emotions are an integral part of interpersonal communication and creating connections between humans. If the mode of communication does not support the transfer of emotions through facial expressions or tone of voice, like in text messages, users often compensate by artificially communicating their emotions, for example, by adding emojis to text messages~\cite{cherbonnier2022recognition}.
In the current landscape of Mixed Reality (MR), and especially Virtual Reality (VR), true emotions are hard to transfer onto virtual avatars, due their fidelity hindering the replicability of subtle or ambiguous 
facial expressions.

Therefore, many techniques utilize emotion recognition~\cite{kakarla2014real, bellenger2024facial}, where emotions are characterized and recognized through face tracking, and can then be used to trigger pre-determined facial animations in avatars. 
This is also especially interesting for accessibility, as emotion recognition is often utilized for systems supporting and understanding the limited emotion recognition skills of neurodiverse people, e.g., with autism spectrum disorder (ASD)~\cite{keating2020facial}. 
Here, several AR approaches utilize emotion recognition to display a simplified emoji to users~\cite{haber2020making,lee2020aegis} or replace faces of others with 3D emojis~\cite{sun2019anonemoji}, 
assisting users in their perception of emotions. 

To use these techniques successfully, the underlying emotion recognition must be as accurate as possible. 
AI-based face tracking is often used for such emotion recognition, as it is widely available and only needs the visual information of the face~\cite{ko2018brief}. 
However, this is still error-prone, as makeup or facial hair can hinder recognition, and facial emotion expressions differ across cultures and individuals (especially those with ASD, who show emotions less frequently, and their expressions are judged by neurotypical people to be less accurate and of lower quality~\cite{trevisan2018facial,loveland1994imitation}). 
Also, artificially over-acted emotions used for training can look different from naturally evoked emotions, especially with regard to intensity. 
Lastly, many recognition systems use distinct, simple categories for labeling, but emotions do not necessarily fit in these categories, as they can be a more complex combination of multiple emotions~\cite{wollmer2008abandoning}.

Thus, we want to discuss several directions to increase inclusivity and improve emotion recognition through facial tracking, either directly on the system's side or combined with user input, as the latter has not yet been deeply explored in the literature.

\section{Improving Inclusivity}
On the system's side, a first step in increasing the quality and inclusivity of emotion recognition is to create more versatile and inclusive emotion datasets to train AI models. Datasets should include acted and naturally evoked emotions, broader scales of emotion intensities, and complex emotion combinations. Additionally, datasets should consist of diverse selections of humans, especially neurodiverse people, as the latter often express emotions differently than neurotypical people, being described as unusual or mechanical-looking expressions~\cite{loveland1994imitation}. 

Without expanding the needed sensors, the system can also use available situational and group contexts in emotion recognition. This means that the system could, for example, use the emotional state of other users, audio from voice chats, or textual information as additional input. 
To make the system more transparent to the user, the results of the emotion recognition should not only be transferred or used as input to trigger functionality but also be communicated to the user. By mirroring to them how the system would interpret their emotional state, they can identify if that matches their true emotions.

Further including the user, static individual calibration of the emotion recognition can be done at the beginning of the tracking, further refined by dynamic calibration through input-driven reinforcement, where the user can tell the system if it recognized the emotion correctly. 
Additionally, we propose that the user can scale the intensity of the recognized emotion to counteract low expressivity or even manually trigger certain emotions if the system cannot recognize them independently.

\section{Discussion \& Conclusion}
In the context of emotions, privacy is an especially important consideration. 
Thus, systems should support user agency by letting them control what personal data can be used to what extent for recognition, and they should be able to turn off the recognition when they do not want to share this information with others. 

Even though we believe that expanding emotion training sets is crucial for making facial emotion recognition systems more inclusive, it is unclear how natural emotions, especially negative ones, can be captured ethically. 
To some degree, emotions could be triggered through external stimuli in a controlled and consensual setting, but the user's knowledge of being recorded can still influence the expression. 
One approach would be to collect self-labeled artifacts created in natural settings (e.g., videos from family gatherings) from volunteers, but this has a significant overhead and might not result in enough high-quality data. 

We believe that systems should not encourage or force users to change their way of expressing emotions, as there is no correct way to do so, and especially in people with ASD, camouflaging and masking (changing behavior to hide ASD related traits) is linked to exhaustion and threats to self-perception~\cite{hull2017putting}. Instead, systems should adapt their recognition to the user and provide them with the tools to help the system understand them, without changing who they are.



\begin{acks}
Funded by the German Research Foundation (DFG, Deutsche For-schungsgemeinschaft) as part of Germany’s Excellence Strategy – EXC 2050/1 – Project ID 390696704 – Cluster of Excellence “Centre for Tactile Internet with Human-in-the-Loop” (CeTI) of Technische Universität Dresden.
\end{acks}

\bibliographystyle{ACM-Reference-Format}
\bibliography{sample-base}


\end{document}